\documentclass[final]{ustcstep}

\usepackage{color}
\usepackage{abstract}
\usepackage{times}
\usepackage{multicol}
\usepackage{graphics}
\usepackage[figuresleft]{rotating}
\usepackage{lscape}
\usepackage{bbding}
\usepackage{amssymb}
\usepackage{array}
\usepackage{textcmds}
\usepackage{float}
\usepackage{ulem}
\usepackage[square]{natbib}
\usepackage{amsmath}

\usepackage{url,hyperref,lineno,microtype,subcaption}

\def\gsim{\;\lower4pt\hbox{${\buildrel\displaystyle >\over\sim}$}\;}
\def\lsim{\;\lower4pt\hbox{${\buildrel\displaystyle <\over\sim}$}\;}
\def\grls{\;\lower4pt\hbox{${\buildrel\displaystyle >\over <}$}\;}

\newcommand{\spder}[2]{\frac{\partial ^ 2{#1}}{\partial{#2}^2}}

\newcommand{\arcsecs}{\hbox{$^{\prime\prime}$}}

\newcommand\addr[2]{{\footnotesize \it $^{#1}$#2}\\}

\begin{document}

\title{The Effect Of Cooling On Driven Kink Oscillations Of Coronal Loops} 

\author{C. J. Nelson$^{1,2*}$, A. A. Shukhobodskiy$^{2,3}$, R. Erd\'elyi$^{2,4}$, M. Mathioudakis$^{1}$\\[1pt]
\addr{1}{Astrophysics Research Centre (ARC), School of Mathematics and Physics, Queen’s University, Belfast, Northern Ireland, UK, BT7 1NN}
\addr{2}{Solar Physics and Space Plasma Research Centre (SP2RC), School of Mathematics and Statistics, University of Sheffield, Sheffield, UK, S3 7RH}
\addr{3}{School of Computing, Creative Technologies \& Engineering, Leeds Beckett University, Leeds, UK, LS6 3QS}
\addr{4}{Department of Astronomy, E\"otv\"os Lor\'and University, P\'azm\'any P\'eter s\'et\'any 1/A, H-1117 Budapest, Hungary}
\addr{*}{Corresponding Author, Contact: c.j.nelson@shefffield.ac.uk}}

\twocolumn[
\begin{@twocolumnfalse}

\maketitle

\begin{abstract}
Ever since their detection two decades ago, standing kink oscillations in coronal loops have been extensively studied both observationally and theoretically. Almost all driven coronal loop oscillations (e.g., by flares) are observed to damp through time often with Gaussian or exponential profiles. Intriguingly, however, it has been shown theoretically that the amplitudes of some oscillations could be modified from Gaussian or exponential profiles if cooling is present in the coronal loop systems. Indeed, in some cases the oscillation amplitude can even increase through time. In this article, we analyse a flare-driven coronal loop oscillation observed by the Solar Dynamics Observatory's {\it Atmospheric Imaging Assembly} (SDO/AIA) in order to investigate whether models of cooling can explain the amplitude profile of the oscillation and whether hints of cooling can be found in the intensity evolution of several SDO/AIA filters. During the oscillation of this loop system, the kink mode amplitude appears to differ from a typical Gaussian or exponential profile with some hints being present that the amplitude increases. The application of cooling coronal loop modelling allowed us to estimate the density ratio between the loop and the background plasma, with a ratio of between $2.05$-$2.35$ being returned. Overall, our results indicate that consideration of the thermal evolution of coronal loop systems can allow us to better describe oscillations in these structures and return more accurate estimates of the physical properties of the loops (e.g., density, scale height, magnetic field strength). 

\section*{Keywords} Solar Corona, Coronal Loop Oscillations, Magnetohydrodynamics, Kink Oscillations
\end{abstract}
\end{@twocolumnfalse}
]

\section{Introduction}
\label{Introduction}

Standing kink oscillations were first observed in coronal loops by \citet{Aschwanden99} and \citet{Nakariakov99} using high-resolution imaging data collected by the {\it Transition Region And Coronal Explorer} (TRACE; \citealt{Handy99}). Those authors found that a flare in the local Active Region (AR) caused the magnetic field guide of the coronal loop to shake from side-to-side in a manner analogous to oscillations of a guitar string. One of the most interesting aspects of kink oscillations in coronal loops was their rapid damping profiles, with many examples of flare-driven coronal loop oscillations damping to sub-resolution spatial scales within two or three periods. Such damping has been explained through a number of physical mechanisms, such as resonant absorption (e.g., \citealt{Ruderman02, Goossens02}), phase mixing, and foot-point damping.

Over the past decade, data from the Solar Dynamics Observatory's {\it Atmospheric Imaging Assembly} (SDO/AIA; \citealt{Lemen12}) has been used to perform statistical analyses of coronal loop oscillations (e.g., \citealt{Zimovets15, Goddard16}). Specifically, \citet{Goddard16} analysed the damping profiles of $58$ coronal loops finding that the amplitudes through time of the majority of oscillations could be modelled by exponential or quasi-exponential (close to but not exactly exponential) profiles. More recently, \citet{Pascoe16} suggested that Gaussian profiles may model damping profiles of coronal loops as well, if not better, than exponential profiles in a large number of cases. It was, however, pointed out by \citet{Goddard16} that some events were non-exponential (and also likely non-Gaussian) meaning, for some coronal loop oscillations, additional physics must be employed to explain the amplitude profiles. Interestingly, it has been shown analytically that it is possible that some driven coronal loops could oscillate and damp in a manner not adequately modelled by Gaussian nor exponential profiles through time.

\begin{figure*}
\includegraphics[scale=0.34,trim={1cm 0 0 0}]{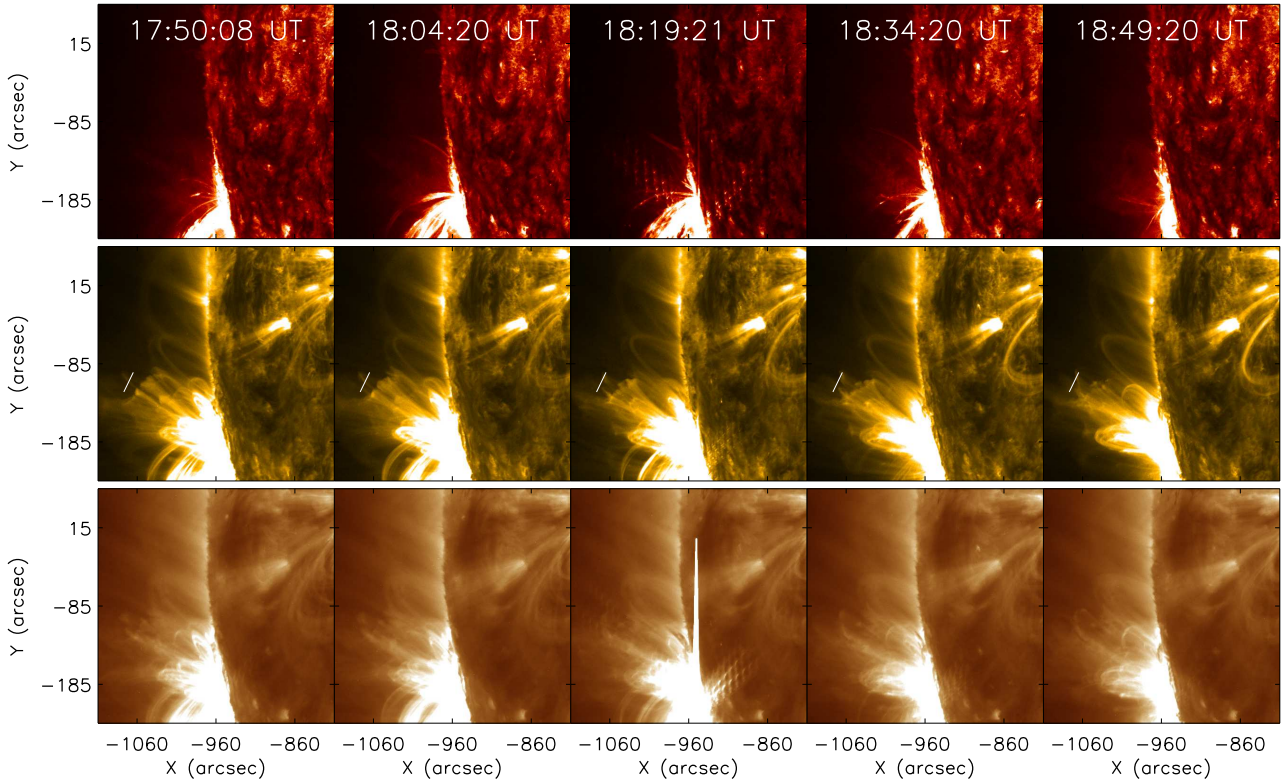}
\caption{The evolution of the coronal loop system during the hour around the oscillations analysed here. Plotted are the SDO/AIA $304$ \AA\ (top row), $171$ \AA\ (middle row), and $193$ \AA\ (bottom row) channels. The white slit over-laid on the middle row indicates the location of the slit analysed in this article.}
\label{Evolution}
\end{figure*}

As the majority of modelling of coronal loop oscillations is conducted using static background parameters (e.g., temperature, density), one opportunity to consider additional physics theoretically is to incorporate some form of time-dependence (see, for example, \citealt{Dymova05, Ghafri13, Erdelyi14}). \citet{Morton09} considered the damping of coronal loops due to cooling through time and found, that for typical oscillatory periods, cooling could play a key role in explaining observed damping profiles (this was shown further in \citealt{Morton10}). The idea that amplification of coronal loop oscillations could occur due to cooling within coronal loops that contained flow was first suggested by \citet{Ruderman11a}. This work was expanded upon in \citet{Ruderman11b} with the inclusion of a resonant layer, where it was found that cooling could cancel out the damping due to resonant absorption in some cases. The effects of loop expansion were considered for cooling coronal loops when no resonant layer was observed by \citet{Ruderman17}. Again it was found that amplification of the coronal loop oscillations could occur.

A more complete analysis comprising of loop expansion, resonant absorption, and cooling was recently conducted by \citet{Shukhobodskiy18a} and \citet{Shukhobodskiy18b} for time independent and time dependent densities, respectively. Those authors found that in some cases, the combined effects of expansion and cooling could dominate over resonant absorption leading to a slowing down of the damping of the oscillations or even, in some cases, a brief amplification in kink mode coronal loop oscillations. Such amplification would only be short-lived and would be followed by the continued decay of the oscillation to sub-resolution levels. Additionally, assuming that both the external and internal densities were longitudinally stratified, the authors showed that the ratio of the frequency of a fundamental mode on the decrement of the kink oscillation is independent on the particular form of the density profile. A similar result was obtained previously by \citet{Dymova06} for kink oscillations in non-expanding magnetic flux tubes. It is possible, therefore, that accounting for cooling when modelling coronal loop oscillations may provide better fits for coronal loop oscillations, especially when increases in amplitude are evident. Additionally, important seismological information could be obtained from the system (such as the density ratio between the loop and the background) if one were to consider cooling which could help improve future numerical modelling. 

The development of modelling of time-dependent coronal loop temperatures has a strong foundation in observations. Numerous authors have discussed the temperature evolution of coronal loop arcades, with cooling often being inferred through the observed progression of loops from hot to cold channels through time (see, for example, \citealt{Winebarger05, Lopez07, Aschwanden08}). It is well known that coronal loops can cool quickly, over the course of two or three oscillatory periods (e.g., \citealt{Aschwanden08}), through processes such as the thermal instability, which can often lead to the occurrence of coronal rain (see \citealt{Antolin15} and references within). Such cooling means the application of theories which are not magnetohydrostatic, such as those recently developed by \citet{Shukhobodskiy18b}, are important for further understanding and modelling coronal loop oscillations more generally. It should be noted that \citet{Morton10} did conduct an application of theoretical work to observed kink oscillations, however, those authors did not include effects which could lead to amplification in their model.

In this article, we apply the theoretical models developed by \citet{Shukhobodskiy18b} to the amplitude profile of a kink oscillation in a coronal loop within AR $11598$. We aim to showcase the seismological potential of cooling models through the inference of the density ratio between the loop and the background plasma. Our work is setup as follows: In Sect.~\ref{Observations} we describe the data studied here and the data analysis methods; In Sect.~\ref{Results} we present our results including the application of the theoretical modelling and {\color{red} any} evidence for cooling in the {\color{red} AR}; In Sect.~\ref{Conclusions} we draw our conclusions and provide some suggestions about future work.

\begin{figure*}
\includegraphics[scale=0.36,trim={2cm 0 0 0}]{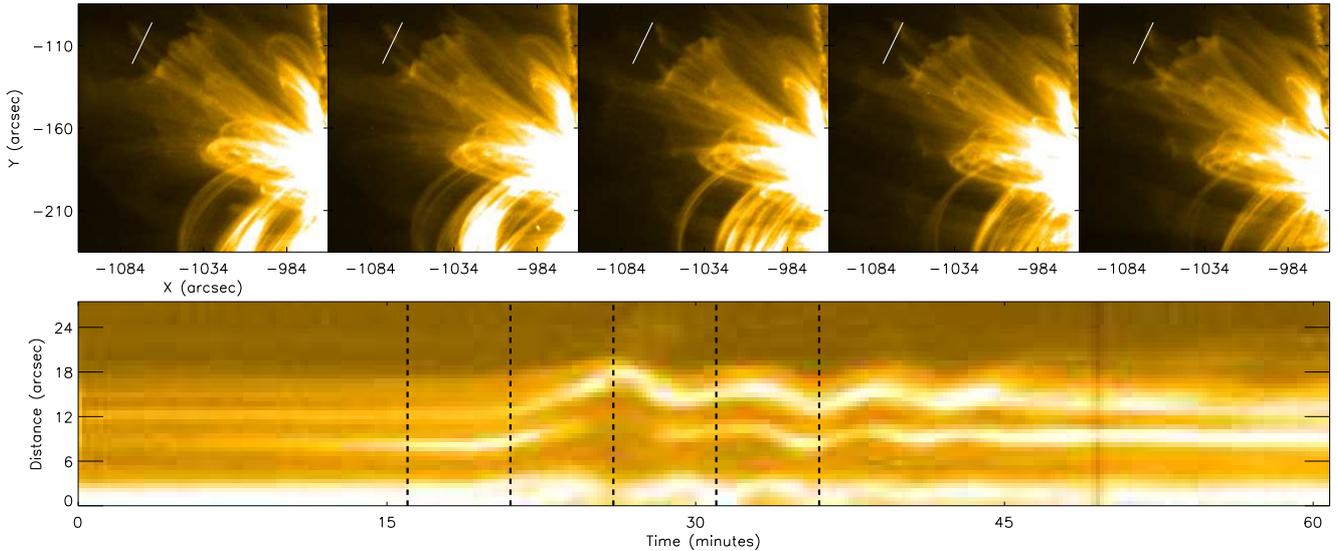}
\caption{(Top row) The coronal loop system analysed here plotted using $171$ \AA\ data at five times during the oscillation. The white lines indicate the slit studied in this article, which corresponds to Loop 40, Event 2 from \citet{Goddard16}. (Bottom row) Time-space diagram constructed from the slit over-laid on the $171$ \AA\ data. The dashed vertical lines indicates the time steps plotted in the top row (from left to right). The bright chord starting and finishing at $12$\arcsecs\ is the oscillation studied in this article.}
\label{Lightcurve}
\end{figure*}

\section{Observations}
\label{Observations}

\subsection{Data And Feature Selection}

The data studied in this article were sampled by the SDO/AIA instrument on the $20$th October $2012$ between $17$:$50$ UT and $18$:$50$ UT. The $300$\arcsecs$\times300$\arcsecs\ field-of-view (FOV) analysed here was focused on AR $11598$ and was initially centred on co-ordinates of $x_\mathrm{c}$=$-959$\arcsecs, $y_\mathrm{c}$=$-85$\arcsecs. Four channels (namely the $304$ \AA, $171$ \AA, $193$ \AA, and $131$ \AA) were downloaded for analysis, using the $ssw\_cutout\_service.pro$ routine, meaning we are able to make inferences about the thermal evolution of the loop system. The loop system is only weakly detectable in the $131$ \AA\ images and, as such, these data are not studied in detail in the remainder of this article. These data do, however, indicate that the coronal loop existed at non-flaring temperatures. Data in the UV sampled by the SDO/AIA instrument have a typical cadence of $12$ s and a pixel scale of $0.6$\arcsecs. As the oscillations analysed here were driven by a flare in the same AR as the coronal loops, some frames were returned with a lower exposure time meaning the loop was difficult to detect due to the lower signal to noise ratio. For these frames, synthetic filling data was created with the intensity of each pixel being taken as the average intensity for that pixel in the frame before and the frame after. This should have no influence on our results as the loop analysed here oscillates with periodicities well above the cadence of that data. This coronal loop oscillation was previously discussed by \citet{Goddard16} (Event 40, Loop 2) and its damping was classified as being best described by a combination of exponential and non-exponential fitting. This event was further analysed by \citet{Pascoe16} and \citet{Pascoe17} who suggested that the presence of multiple harmonics could explain the complicated amplitude profiles observed within this event.

The general evolution of the loop system analysed here is plotted in Fig.~\ref{Evolution} for the $304$ \AA\ (top row), $171$ \AA\ (middle row), and $193$ \AA\ (bottom row) channels. At $17$:$50$ UT, the coronal loop system can be detected in both the $171$ \AA\ and $193$ \AA\ channels, with the white line over-laid on the $171$ \AA\ images indicating the location of the slit studied here. The location of the flare in the AR, which occurred at approximately $18$:$08$ UT, is pin-pointed by the typical disturbance patterns on the $193$ \AA\ image in the third column. In the southern part of the FOV (around $100$\arcsecs\ south of the oscillation analysed here), a large coronal loop arcade can be observed in all panels up until  $18$:$19$ UT. Over the subsequent $30$ minutes (between $18$:$20$ UT and $18$:$50$ UT), however, the loop system completely fades from view in the coronal channels and large amounts of coronal rain are detected in the $304$ \AA\ data, draining material from the loops. The loop system completely fades from view by $18$:$49$ UT.

\subsection{Tracking The Loop Displacement And Model Fitting}

\begin{figure*}
\hspace{0.5cm}
\includegraphics[scale=1.8,trim={0.3cm 0 0 0}]{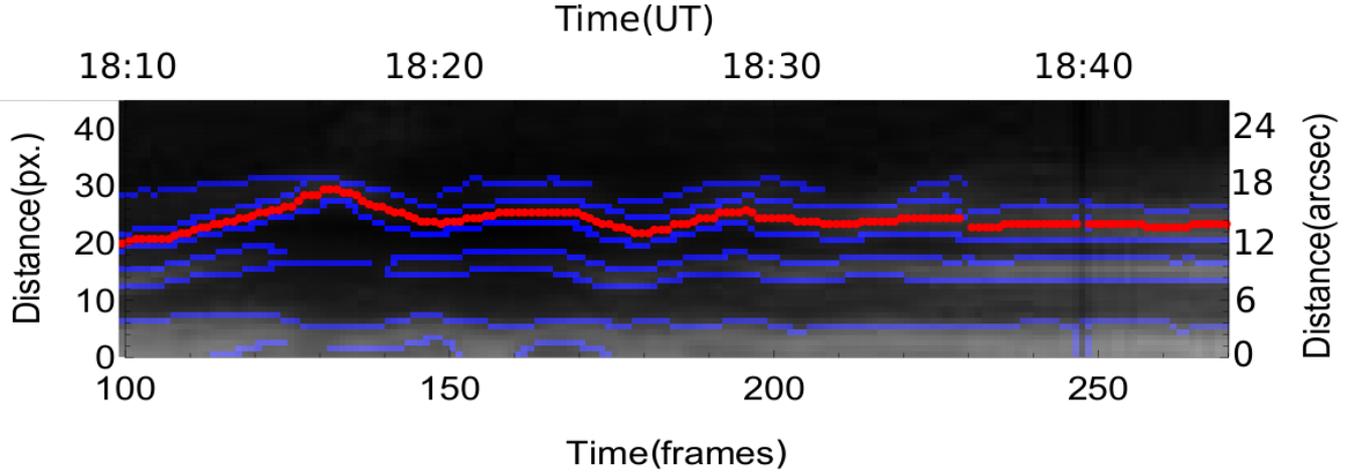}
\caption{Zoomed time-space diagram from Fig.~\ref{Lightcurve} with the Canny edge-detection output (blue lines) and the returned oscillation profile (red line) over-laid. The returned oscillation is qualitatively similar to the oscillation returned by \citet{Pascoe16} for the same event.}
\label{TimeDistanceSlit}
\end{figure*}

In Fig.~\ref{Lightcurve}, we plot a zoomed in FOV of the oscillation analysed here for five time-steps (top row). The white line over-laid on theses images indicates the location of the slit analysed in this article. The co-ordinates of this slit are the same as for Loop 40, Event 2 included in the table of \citet{Goddard16}. This event was studied subsequently by \citet{Pascoe16} and \citet{Pascoe17}. Following the construction of the time-space diagram from $171$ \AA\ images (plotted in the bottom panel of Fig.~\ref{Lightcurve}), we applied a Canny edge-detection algorithm to obtain approximations for the coronal loop boundaries. The output for this edge detection routine is over-laid on the time-space diagram in Fig.~\ref{TimeDistanceSlit} (blue lines). In order to model the oscillation of the loop axis, we assumed that the displacement is guided by midpoints between the boundaries of the upper loop as shown by the red dots in Figure \ref{TimeDistanceSlit}. If we assume that the loop is cylindrical and radially symmetric, then the errors associated by this fitting can be assumed to be less than $1$ pixel. Understanding whether such assumptions are justified would require much higher resolution data and so should be studied in the future. As the loop axis returned through time is qualitatively similar to that plotted in Fig.1 of \citet{Pascoe16} for the same event, in the following we assume that this oscillation profile is accurate and, therefore, neglect any further potential errors in our fitting. Ignoring any further errors introduced through the loop fitting is justified in this case, as our study aims to show, in principle, that including the effects of cooling could have important implications for coronal loop modelling. A larger statistical study conducted using a variety of fitting methods in the future will be required to fully understand the prevalence and importance of coronal loop cooling on kink-mode oscillations.

\begin{figure*}
\hspace{4.8cm}
\includegraphics[scale=1, trim={1cm 0 0 0}]{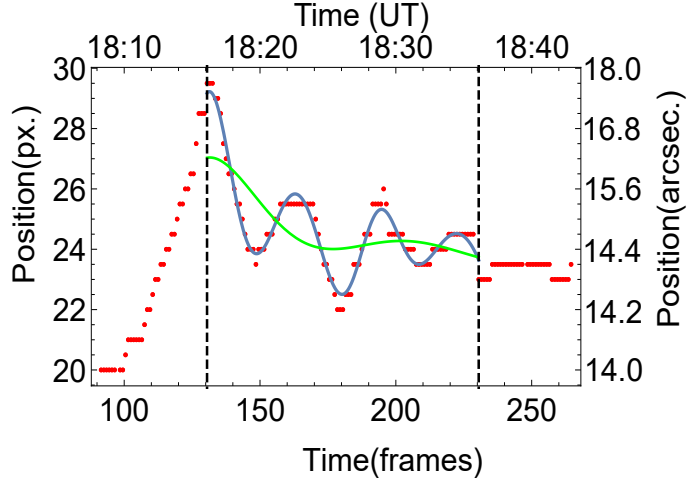}
\caption{The position-time dependence of the coronal loop. Red dots correspond to data points collected from the $171$ \AA\ time-space diagram plotted in Fig.~\ref{TimeDistanceSlit}. The green line corresponds to the fitted $8$th-order background trend and the blue line is the Gaussian fit of the observed data points computed of the form Eq.~\ref{Gaussian}. The two vertical dashed lines indicate the boundaries of the area over which numerical fitting took place. The Gaussian fit under-estimates the oscillation amplitude between frames 170 and 200 meaning the amplitude at this time will be under-estimated.}
\label{GauFit}
\end{figure*}

The oscillation extracted from this time-space diagram is plotted with red dots in Fig.~\ref{GauFit}. The approximate background trend of the loop was modelled by a polynomial of the 8th degree on all obtained data points (i.e., from frame $0$ to frame $300$) and is over-laid for the region of interest (between frames $130$ and $230$) as the green line in Fig.~\ref{GauFit}. The $8$th order polynomial was selected as it best tracked the background of the amplitude profile throughout the entire time-series (including the parts where no oscillation was detected). It should be noted that a number of background trends with different orders were tested with each returning similar seismological results. This is because the ratio of the internal and external densities, which we can calculate from this model, is dependent only on where, in this case, the increment to the amplitude is (which did not change for any background trend), not how large it is. This will be discussed in further detail later in the article. 

The numerical fitting of the red points is done by considering the summation of four equations of Gaussian form: 
\begin{equation}
f(amp)=\sum_{i=1}^{4} A_{i} \exp[-(-\mu_{i}+t)^2 /(2 \sigma_{i}^2)]/(\sqrt{2 \pi}/\sigma_{i}^2),
\label{Gaussian}
\end{equation}
where $A_{i}$\/, $\mu_{i}$\/ and $\sigma_{i}$ are variables to be fitted for each peak, $t$ is time (in frames), and $f(amp)$ is the final fitted function. The benefit of such fits over a typical sinusoidal fit is that no periodicity is assumed a priori. This fitting was completed using the Wolfram Mathematica $11.3$ procedure {\it NonlinearModelFit} (which guarantees continuity of the summed functions) and is plotted as the blue line over-laid on Fig.~\ref{GauFit}. It is clear that this blue line accurately models the detected oscillation, however, it does slightly under-estimate the returned amplitude between frames 170 and 200. Therefore, we stress that the increase in amplitude detected during this oscillation (discussed subsequently) will be under-estimated by our model.

\section{Results}
\label{Results}

\subsection{Theoretical Modelling And Observed Amplitude Profiles}

\begin{figure*}
\hspace{4.cm}
\includegraphics[scale=0.8]{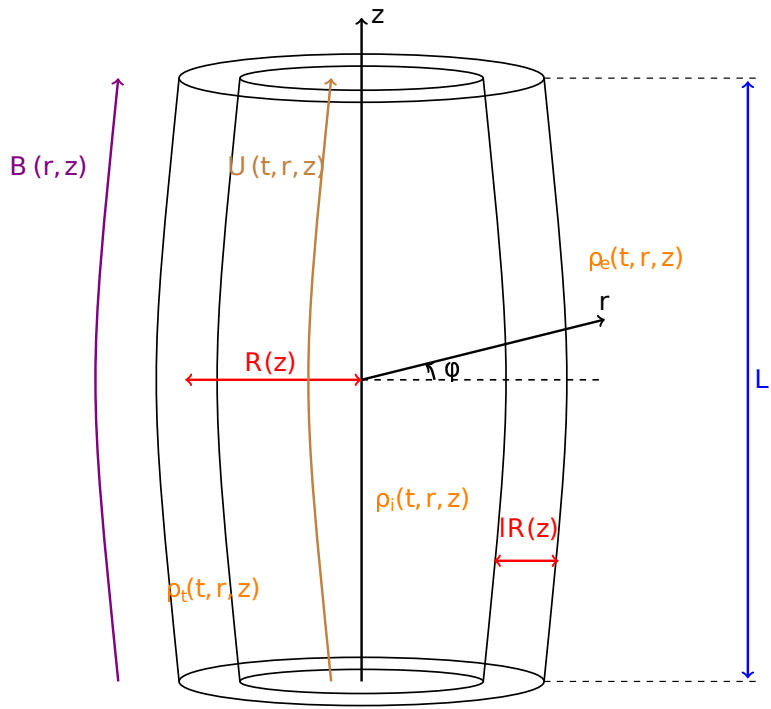}
\caption{Equilibrium configuration of theoretical model. The black lines with labels $r$\/, $\phi$ and $z$ correspond to the cylindrical polar coordinates system. $\bold{B}(r,z)$ is the background magnetic field, $\bold{U}(t,r,z)$ is the background flow parallel to the magnetic field, $R(z)$ is the radius of the tube, $l$ is a constant determining the thickness of the transitional layer, $\rho_i(t,r,z)$ is the density in the internal region, $\rho_e(t,r,z)$ is the density in external region, $\rho_t(t,r,z)$ is the monotonically decaying density in the transitional layer from $\rho_i(t,r,z)$ to $\rho_e(t,r,z)$, $L$ is the length of tube for the standing waves and the characteristic value of the wavelength for propagating waves. }
\label{EQCON}
\end{figure*}

In this section, we fit the observed oscillatory profile plotted in Fig.~\ref{GauFit} using the theoretical model proposed by \citet{Shukhobodskiy18b}. The model proposed by those authors consisted of a straight magnetic flux tube with varying cross section along its length. The tube was modelled with three layers: the core which contained an arbitrary flow and oscillated as a solid body; a transitional layer (or annulus) with monotonically decaying density from the core to the external layer; and the surrounding background plasma. See Fig.~\ref{EQCON} for a detailed schematic of this model. \cite{Ruderman17} obtained the governing equation for such a model under the assumptions of a thin tube with thin boundaries, by considering jumps of displacement and pressure across the transitional layer. However, this equation was not closed as jump conditions were not defined in terms of the displacement. Later, \citet{Shukhobodskiy18a} and \citet{Shukhobodskiy18b}, closed that system for the time independent and time dependent densities, respectively. 

In the model applied here, we consider a loop of half-circular shape, where the curvature affects only the density distribution. The temperature is modelled to be exponentially decaying in time (similar to the profiles studied in, for example, \citet{Aschwanden08, Morton10, Ruderman11b, Ruderman17}) and is approximated by:
\begin{equation}
 T(t)=T_0 \exp(-t/t_{cool}),
 \label{eq:3.3.1}
\end{equation}
where $T_0$ is the constant external temperature and $t_{cool}$ is the cooling time (assumed to be the total lifetime of the oscillation here). Additionally, we describe the variation of the loop cross-section with height, $z$, in a manner similar to that discussed by \citealt{Ruderman08, Ruderman17, Shukhobodskiy18b}. This takes the form:
\begin{equation}
R(z) = R_f \lambda \sqrt{\frac{\cosh(L/2L_c)-1}{\cosh(L/2L_c) - 
  \lambda^2+(\lambda^2-1)\cosh(z/L_c)}},
\label{eq:3.3.2}
\end{equation}
where $R_f$ is the radius of the magnetic flux tube at the foot-points, $\lambda=R(0)/R_f$ is the expansion factor and $L_c$ is an arbitrary constant. Please note that $L_c$ should be selected such that the expansion factor can be consistent with observations of coronal loops ($1<\lambda<1.5$). We also assume, similarly to \citealt{Ruderman02, Goossens02}, that the density in the transitional layer has a linear profile modelled by:
\begin{equation}
  \rho_t(r,z) = \frac{\rho_i + \rho_e}2 + (\rho_i - \rho_e)\frac{R - r}{lR},
\label{eq:3.3.3}
\end{equation}
where $l$ is a constant determining the width of the transition layer and $\rho_\mathrm{i}$ and $\rho_\mathrm{e}$ are the internal and external densities, respectively. 

Using these equations, \citet{Shukhobodskiy18b} showed that the dimensionless amplitude, $A$ where $A(0)=1$, of the kink mode may be approximated by:
\begin{equation}
  \frac{d (\varpi \Pi_{+} A^2)}{dt} = - \alpha \varpi^2 |\Pi_-| A^2,
\label{eq:3.3.4}
\end{equation}
where:
\begin{equation}
  \Pi_\pm = \int^1_0 X^2 \lambda^4 \left[\zeta\exp\left(-\kappa e^{\tau} 
  \cos\frac{\pi z}2\right) \pm \exp\left(-\kappa \cos\frac{\pi z}2\right)\right] dZ,
\label{eq:3.3.5}
\end{equation}
\begin{equation}
 \alpha = \frac{\pi l C_f t_{cool}}{4 L},
\label{eq:3.3.6}
\end{equation}
and:
\begin{equation*}
 \zeta=\frac{\rho_i}{\rho_e}, \  Z = \frac{2 z}{L}, \ \tau = \frac{t}{t_{cool}}, \ 
  \varpi = \frac{\omega L}{C_f}, 
\end{equation*}
\begin{equation}
 \kappa = \frac{L}{\pi H_0}, \
C_f^2 = \frac{2\zeta B_f^2}{\mu_0\rho_f(1 + \zeta)}.
\end{equation}
Here, $\mu_0$ is the magnetic permeability of the free space, $\omega$ is the oscillation frequency, $H_0$ is the external scale height, and $\rho_f$ and $B_f$ are the density of plasma and magnetic field strength at the foot-points inside the loop. Additionally, $X$ is determined by the following boundary value problem:
\begin{equation*}
  \spder{X}{Z} +
\end{equation*} 
\begin{equation}\frac{\varpi^2 \lambda^4X}
  {4 (\zeta +1)} 
  \left[\zeta\exp\left(-\kappa e^{\tau} \cos\frac{\pi Z}2\right) + 
  \exp\left(-\kappa \cos\frac{\pi Z}2\right)\right] = 0,
\label{eq:3.3.7}  
\end{equation}
where $X$ is the function of $Z$ defined by the boundary values:
\begin{equation}
  X = 0 \quad \text{at} \;\; Z = -1, \quad X = 0 
  \quad \text{at} \; \;Z = 1.
\label{eq:3.3.8}
\end{equation}

\begin{figure*}
\hspace{3.4cm}
\includegraphics[scale=1]{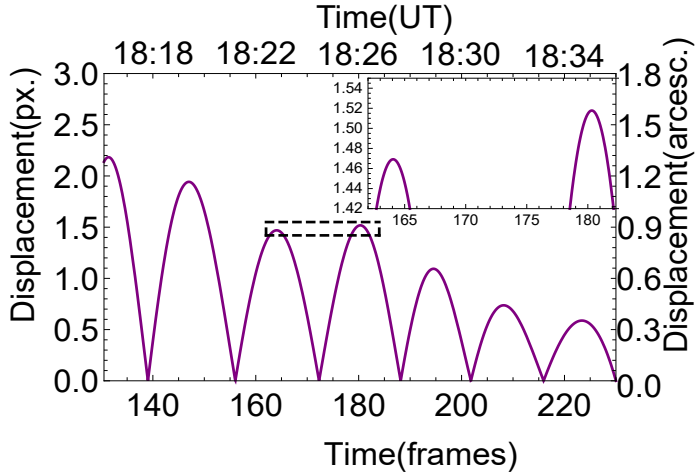}
\caption{The absolute value of the displacement of the Gaussian fitted position profile (blue line from Fig.~\ref{GauFit}) from the background $8th$ order polynomial trend (green line from Fig.~\ref{GauFit}) through time. The cut-out in the top right corner provides a magnified view of the region on the graph bounded by the black dashed box. This cut-out clearly shows the deviation of the damping profile from a typical Gaussian or exponential fit during this time due to a slight increase in the measured amplitude (compared to the background trend). Similar results are found for other background polynomial trends.}
\label{AbsDis}
\end{figure*}

In order to obtain results comparable with observed amplitude profiles, we set $A_t = A(0) A_{Ob}(0)$\/, where $A_{Ob}(0)$ is the measured initial amplitude of the observed oscillation, $A_t$ is the factor by which the dimensionless amplitude is scaled, and set $L/L_c=6$ (similar to the values used by \citealt{Ruderman08, Ruderman17, Shukhobodskiy18b}). This allows us to obtain results for loop expansion factors in the range of $1$--$1.5$, consistent with values discussed in the literature for coronal loops (see, for example, \citealt{Klimchuk00, Watko00}). In Fig.~\ref{AbsDis}, we plot the absolute displacement of the observed oscillation through time by subtracting the fitted displacement (blue line in Fig.~\ref{GauFit}) from the background trend (green line in Fig.~\ref{GauFit}). We define the initial amplitude of the oscillation, $A_{Ob}(0)$, by the first peak in Fig.~\ref{AbsDis}. The observed amplitude profile is then constructed by calculating all values of the peaks in Fig.~\ref{AbsDis} and using the {\it Interpolation} procedure in Wolfram Mathematica 11.3.

The most interesting region of Fig.~\ref{AbsDis} (denoted by the dashed black box over-laid between frames 160-185) is plotted in the zoomed cut-out in the top right corner of the panel. It is immediately clear that peak four is slightly larger than peak three indicating an amplitude increase potentially occurs through time. Such an increase would be in agreement with the theoretical model of \citet{Shukhobodskiy18b} which suggests that cooling in the loop system can lead to wave amplification. We note that cooling does not strictly lead to an amplification of the oscillation, it only modifies the damping profile from Gaussian or exponential. Therefore, although the amplitude increase plotted here  is sub-resolution, the fact that the oscillation is not damping in a manner consistent with a Gaussian or exponential profile is enough to suggest some sort of cooling may be occurring. It should also be noted that larger amplitude increases (greater than 1 pixel) were found for lower order polynomials, however, we only focus on the $8th$ order fitting here as that is sufficient as a proof of concept. The increase in the amplitude can be clearly seen in Fig.~\ref{TheorObs}, where the blue line plots a fit between the measured peak amplitudes (green circles) through time. The red line over-laid on Fig.~\ref{TheorObs} indicates the line of best fit calculated by solving the system of equations (\ref{eq:3.3.1})--(\ref{eq:3.3.8}) numerically. A maximum $\chi^2$ value was found by looping over all variables, hence this is numerically expensive. It was assumed that the cooling starts at the first peak and ends in the last peak of Fig.~\ref{AbsDis} (i.e., that $t_{cool}$ is equal to the lifetime of the oscillation). 

The software used to obtain these theoretical solution is Wolfram Mathematica $11.3$. The numerical procedure for obtaining the theoretical results may be summarised as follows: To obtain the solution to the boundary value problem, Eqs.~(\ref{eq:3.3.7}) and (\ref{eq:3.3.8}), we used the {\it NDEigensystem} procedure; Eq.\~ (\ref{eq:3.3.5}) was then integrated numerically using the {\it NIntegrate} procedure subject to the {\it GlobalAdaptive} method (which uses various numerical integration methods and chooses the most accurate and fastest version for the particular problem). Finally we obtained the amplitude of the oscillation by using the {\it NDSolve} procedure.

The application of this model revealed several interesting effects with regards seismology of the loop system. Firstly, it was found that $\zeta$ (the density ratio at the foot-points of the loop at $t=0$) and $\kappa$ (the scale height) determine the position of the turning points of the amplitude profile (i.e., where the gradient of the amplitude profile is zero). However, the effect of $\kappa$ on the position of the turning point reduces as the value of $\kappa$ itself increases. As a result, we can neglect this effect for sufficiently long loops with $\kappa>1.6$. Therefore, by minimising the difference between the turning points of the theoretical and observed amplitude profiles we can determine an approximate value of $\zeta$ for this system without knowing any other background parameters. For the example studied here, fitting a density ratio of between $2.05$ and $2.35$ provides sufficiently good approximations for the position of the local amplitude increase. As this work aims to provide a proof-of-concept of the application of cooling theory to coronal loop oscillations and as the density ratio is only dependent on where the amplitude deviates from a Gaussian or exponential decay, we neglect any errors in the measured amplitude (green dots) here; however, the effects of any errors on our seismological should be analysed using a larger statistical sample in the future. The differences in nature between the red and blue curves are due to the spline fitting used here. Secondly, the values of $\kappa$\/, $\alpha$, and $\lambda$ determine the speed of the amplitude decay. Unfortunately, for the example studied here, various sets of these parameters return similar shapes and amplitude profiles meaning we are unable to provide accurate seismological estimates at this stage. A larger statistical sample of cooling events and further numerical work will be required to attempt to estimate such values in the future.

\subsection{Evidence For Cooling In The AR}

Evidence of cooling in the coronal loop system can be inferred through analysis of the intensities within SDO/AIA imaging channels. For the loop studied here, no evidence of coronal rain formation was found in the $304$ \AA\ filter indicating that catastrophic cooling likely did not take place within this loop. The loop intensity within the $171$ \AA\ and $193$ \AA\ filters did appear to decrease slightly (potentially below any level of significance) during the oscillation, however, whether this was due to cooling, some line-of-sight effects (e.g., the supposition of multiple loops within one pixel), or purely noise is unknown. The flare within the local AR lead to large swathes of saturation and fringing patterns on the $193$ \AA\ data meaning we were unable to study the temporal evolution of the loop in detail in this wavelength during the period of interest for cooling for this event. Future work should aim to analyse a larger statistical sample of coronal loop oscillations with non-Gaussian and non-exponential damping profiles in order to detect whether direct signatures of cooling can be found within the loops themselves.

Significant cooling can be inferred elsewhere in this AR during the course of this oscillation, however. Large amounts of coronal rain can be observed in the $304$ \AA\ channel in the loop system approximately $100$\arcsecs\ south of the loop analysed here potentially indicating the occurrence of the thermal instability during the flare. The after-effects of this rain are evident in the top row of Fig.~\ref{Evolution} (at approximately $x_\mathrm{c}$=$-200$\arcsecs, $y_\mathrm{c}$=$-990$\arcsecs). Initially (at $17$:$50$ UT), the loop is bright in the $304$ \AA\ channel. As the rain forms, the loop system completely fades from view (by $18$:$50$ UT). The coronal rain in the chromospheric $304$ \AA\ channel is accompanied by a reduction in the intensity of the loop system in the coronal $171$ \AA\ and $193$ \AA\ channels further supporting the hypothesis that cooling occurred. The application of the theories tested here on loops which are catastrophically cooling would be an interesting project for the future.

\section{Discussion And Conclusions}
\label{Conclusions}

In this article, we have studied a flare-driven kink oscillation in a coronal loop observed in AR $11598$ by the SDO/AIA instrument (Fig.~\ref{Evolution}). This oscillation corresponded to Loop 40, Event 2 from \citet{Goddard16}. The specific slit analysed here is indicated by the white line over-laid on the top row of Fig.~\ref{Lightcurve}, with the returned time-space diagram being plotted in the bottom panel of Fig.~\ref{Lightcurve}. In Fig.~\ref{TimeDistanceSlit} the time-space diagram is plotted with the output from a Canny edge-detection algorithm over-laid in blue. The mid-points between the edge-detection outputs (red dots) were assumed to track the displacement of the centre of the loop as it oscillated. The returned oscillation was qualitatively similar to that returned by \citet{Pascoe16} for the same event indicating that our fitting method was sound. This oscillation was then fitted with a summation of four Gaussian functions (corresponding to the four peaks in the oscillation) of the form Eq.~\ref{Gaussian} and an $8$-th order background trend removed (the blue and green lines in Fig.~\ref{GauFit}, respectively). Again, it should be emphasised that numerous background trends were tested in order to assure that we were not introducing important effects in our subsequent analysis.

Once the oscillation and the background had been fitted, we removed the background trend from the oscillation and plotted the absolute values of the kink mode (see Fig.~\ref{AbsDis}) through time. The cut-out over-laid on Fig.~\ref{AbsDis} clearly highlights a deviation in oscillatory amplitude from a typical Gaussian or exponential damping between frames $160$ and $185$ ($18$:$22$ UT - $18$:$27$ UT). Indeed, there are even hints that an amplitude increase could be present, however, this is small. It should be noted that some background trends returned profiles with amplitude increases of over 1 pixel. Such an amplitude increase would obviously not be expected from a typical Gaussian or exponential decay (as has been considered previously by, for example, \citealt{Goddard16, Pascoe16}). However, models proposed by \citet{Shukhobodskiy18a} and \citet{Shukhobodskiy18b} which consider cooling within the system can account for amplitude increases. Therefore, in addition to the multiple harmonics scenario suggested by \citet{Pascoe17}, future work should consider cooling when trying to understand complex amplitude profiles in coronal loop oscillations.

\begin{figure*}
\hspace{3.4cm}
\includegraphics[scale=1]{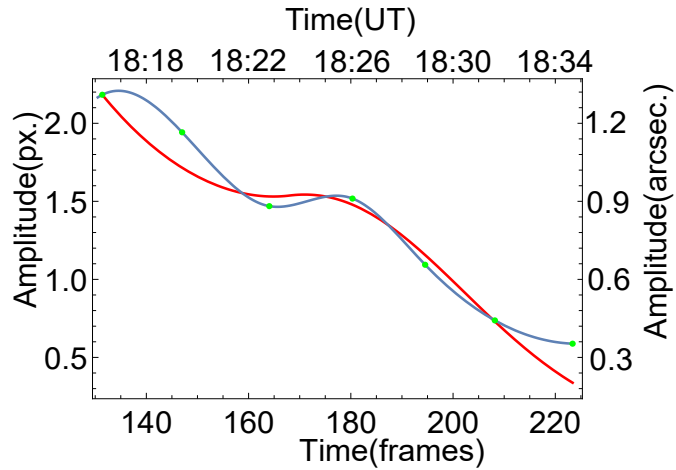}
\caption{The dependence of the amplitude of the coronal loop oscillation on time. Green points correspond to maxima of the peaks plotted in Fig.~\ref{AbsDis}, the blue line is the fitted observed amplitude-time profile, and the red line represents the theoretical amplitude-time profile with $\zeta=2.2$\/, $\lambda=1.05$\/, $\kappa=2.4$ and $\alpha=1.4$\/.}
\label{TheorObs}
\end{figure*}

Although there were some hints that the loop faded slightly in the coronal $171$ \AA\ and $193$ \AA\ channels over the course of the hour-long dataset analysed here, it is unclear whether this is an effect of cooling, due to line-of-sight effects, or purely noise. Little other evidence for cooling could be obtained through analysis of SDO/AIA time-series. Additionally, no evidence of coronal rain was observed in the $304$ \AA\ channel. More obvious cooling could be observed within the local AR, however, in a loop system located approximately $100$\arcsecs\ south of the loop analysed here. This loop contains large amounts of coronal rain in the $304$ \AA\ channel potentially hinting at the occurrence of the thermal instability, or catastrophic cooling. The effects of such catastrophic cooling on coronal loop oscillations would be an interesting topic for future study.

The key seismological result obtained here is that the location in time at which the amplitude begins to increase (an effect of the cooling on the oscillation) is dictated solely by the ratio between the internal and external densities for sufficiently large coronal loops. Therefore, fitting the model such that the difference between the observed turning point and the theoretical fitting point is minimised (as is shown in Fig.~\ref{TheorObs}) allows us to calculate an estimate of the density ratio, $\zeta$, an important parameter for modelling. For this loop, inversions of the theoretical model provide a good fit to the observed amplitude profiles when $\zeta$ is in the range $2.05$--$2.35$ (i.e., the loop foot-point is initially more than twice as dense as its surroundings).

In order to follow on from this work, we aim to complete two further studies. Firstly, we will conduct a statistical analysis of oscillations within potentially cooling coronal loops in the solar corona. This will provide important constraints on density ratios of loops for future modelling. Secondly, through further numerical work, it should prove possible to conduct further seismology in order to return values such as the scale height of the loop and the annulus thickness. These values will, again, provide further constraints for future modelling. Overall, the theoretical work of \citet{Shukhobodskiy18b} has proved adept at modelling the oscillations of the loop analysed here and should be considered by authors in the future when analysing coronal loop oscillations.

\section*{Conflict of Interest Statement}

The authors declare that the research was conducted in the absence of any commercial or financial relationships that could be construed as a potential conflict of interest.

\section*{Author Contributions}

CN drafted the manuscript and lead the observational analysis with help from AS. AS drafted Section 3.1 and completed the theoretical work. AS and CN completed the model fitting. All authors contributed to the interpretation of the results and helped draft the manuscript.

\section*{Funding}

The authors acknowledge the Science and Technology Facilities Council (STFC) for financial support (Grant numbers: ST/M000826/1 and ST/P000304/1).

\section*{Acknowledgments}
SDO data are courtesy of NASA/SDO and the AIA and HMI science teams.

\section*{Data Availability Statement}
The datasets analyzed for this study can be downloaded using standard routines in IDL and Python from open-source SDO data repositories using the information included in the Observations section.

\bibliographystyle{frontiersinHLTH_FPHY}
\bibliography{Cooling_Coronal_Loops}

\end{document}